# Análise comparativa de pesquisas de origens e destinos: uma abordagem baseada em Redes Complexas

**Charles Marques, Carlos Caminha[1], Vasco Furtado[1]**

[1]Programa de Pós-graduação em Informática Aplicada – PPGIA da Universidade de Fortaleza (Unifor) – Campus Unifor

charles.marques[@]outlook.com, caminha[@]unifor.br, vasco[@]unifor.br

*Abstract.*

*In this paper, a comparative study was conducted between complex networks representing origin and destination survey data. Similarities were found between the characteristics of the networks of Brazilian cities with networks of foreign cities. Power laws were found in the distributions of edge weights and this scale - free behavior can occur due to the economic characteristics of the cities.*

*Resumo.*

*Neste trabalho foi conduzido um estudo comparativo entre redes complexas que representam dados de pesquisa de origem e destino. Foram encontradas similaridades entre as características das redes de cidades brasileiras com redes de cidades estrangeiras. Foram encontradas leis de potência nas distribuições dos pesos das arestas e esse comportamento livre de escala pode ocorrer devido a características econômicas das cidades.*

## 1. Introdução

Compreender padrões de mobilidade humana em grandes cidades sempre despertou interesse da ciência [GONZALES, HIDALGO, BARABASI, 2008], [BARABASI ET AL, 2010], [CHONG ET AL, 2011]. Um instrumento tradicional para esse fim são as pesquisas qualitativas sobre origem e destino (OD) [CAMINHA ET AL, 2016]. Nelas são realizadas entrevistas que buscam capturar amostras de intenções de viagens de forma que sejam representativas da movimentação total das pessoas [ABRAHAMSSON, 1998].

Alternativamente a esses métodos qualitativos vêm surgindo "pesquisas" OD feitas quantitativamente a partir do rastreamento de movimentos das pessoas. Consegue-se amostras normalmente bem superiores àquelas feitas pelos métodos de entrevista e que não capturam somente a intenção de viagem das pessoas, mas a própria viagem realizada. Em função de serem menos onerosas e rápidas de serem realizadas, a quantidade de pesquisas OD quantitativas em grandes cidades tem aumentado significativamente.

Esse contexto de incremento de dados sobre movimentação das pessoas requer o desenvolvimento de estudos e métodos que sejam capazes de extrair conhecimento de bases de dados que representam amostras, capturadas qualitativa ou quantitativamente. Redes complexas se apresentam como formalismo adequado para apoiar esse processo de descoberta, pois são capazes de revelar padrões ou assinaturas comparáveis entre dados de diferentes cidades e capturados de forma diferente.

Um exemplo desse tipo de estudo é o trabalho feito por Saberi (2016), que modelou duas redes a partir de dados adquiridos por pesquisas qualitativas de OD de Melbourne e Chicago. Seu estudo concluiu que as redes de OD das duas cidades têm propriedades distintas, resultado das diferentes características comerciais das cidades. Enquanto Melboune tem vários polos comerciais homogeneamente distribuídos pelo seu espaço geográfico, Chicago mostra-se mais concentradora com um grande polo comercial (centro comercial) bem mais destacado em relação a outros pontos da cidade.

Nosso trabalho de pesquisa segue metodologia similar ao de Saberi. Conduzimos um estudo comparativo entre três metrópoles brasileiras e as cidades estrangeiras estudadas por Sabrei (2016). Utilizamos, para isso, dados qualitativos de pesquisas OD de Belo Horizonte e São Paulo, além de dados quantitativos de OD advindos de padrões de uso do sistema de ônibus de Fortaleza.

Observou-se um grupo formado por Fortaleza, Belo Horizonte e Chicago com características que as aproximam em termos de mobilidade e que as distanciam do outro grupo formado por São Paulo e Melbourne. Dentre as características estudadas o que mais chamou atenção foi que Fortaleza, Belo Horizonte e Chicago apresentam uma surpreendente lei de potência na sua distribuição dos pesos das arestas, que no contexto deste trabalho representam a quantidade de viagens realizadas dentro das cidades. Seguindo a mesma conjectura feita por Saberi, acreditamos que as características comerciais das cidades e a forma como elas impactam na ocupação do solo explicam essa similaridade na mobilidade urbana e fazem aparecer essa assinatura própria evidenciada pela lei de potência. A existência de um polo mais desenvolvido que se torna o hub comercial leva a um padrão global de deslocamento com essa propriedade,

## 2. Conjuntos de Dados e Metodologia

Foram utilizados nesse trabalho os conjuntos de dados de viagens de viagens de três cidades brasileiras: Fortaleza, Belo Horizonte e São Paulo, e também os dados estatísticos das cidades Melbourne (Austrália) e Chicago (EUA) obtidos de Saberi (2016). Os dados de origem e destino de viagens de Fortaleza foram obtidos de Caminha (2016) e são referentes ao dia 11 de março de 2015 com um total de 294.869 viagens entre paradas de ônibus com as informações de Lat/Lng das paradas de origem e de destino. Os dados de Belo Horizonte [BHTRANS, 2016] e com 155.292 viagens no ano de 2002 e 140.540 viagens em 2012. Já São Paulo em 2007 [METRÔ 1, 2016] conta com 196.699 viagens com os dados de origem e destino. Já em 2012 [METRÔ 2, 2016] foram 31.854 viagens.

Conforme dados do IBGE 2 (2016) Fortaleza possui 3043 setores censitários, Já Belo Horizonte possui 3936, e São Paulo possui 18.953. Belo Horizonte também está organizada em Área Heterogênea, equivalente ao setor censitário [BHTRANS, 2016], com 1255 áreas heterogêneas.

A pesquisa foi segmentada em cinco etapas: levantamento do estado da arte; obtenção dos dados; uniformização; e transformação e análise de redes complexas. Na primeira etapa foi feito o levantamento do estado da arte, embasamento teórico e aspectos importantes na área. Na segunda etapa, foi realizado uma busca por conjuntos de dados de redes de origem e destino no Brasil e no mundo. Após obter os conjuntos foi necessária uma etapa de uniformização dos dados. Posteriormente à uniformização dos dados, na quarta etapa, eles foram transformados para o modelo de redes complexas. Construídas as redes complexas, iniciou-se à quinta etapa do estudo: a geração e análise das propriedades e distribuições.

Para a transformação e padronização dos dados foram utilizadas as ferramentas: *Java* versão 8, *release* 91. Também a biblioteca *Java Jung*, versão 2.0.1 j [JUNG, 2016], e a *JavaApiForKml* versão 2.2.0 [JAK, 2016] além da *IDE Eclipse Mars* 2 [ECLIPSE, 2016]. Para calcular grande parte das métricas de redes complexas deste trabalho foi utilizado o software *OpenSource GEPHI* [GEPHI, 2016]. Para estimar regressões e visualizar distribuições foi utilizado o software estatístico *QtGrace* [QTGRACE, 2016].

As ferramentas *Java*, *Jung* e *JavaApiForKml* e *Eclipse* foram utilizadas para construção do programa que recebe as informações brutas e as transforma no grafo para então serem realizadas as análises que são objetivo desse trabalho.

O sistema possui uma classe que possui três métodos para conversão dos dados: o primeiro para conversão do arquivo com os dados de viagens em UTM [PEREIRA, 2016] para arquivo de viagens em Latitude/Longitude [SILVA, GUALBERTO, TUPINAMBÁS, 2013]; o segundo para conversão de coordenadas Latitude/Longitude de origem e destino para Setor Censitário, que é a área de trabalho do recenseador [IBGE 1, 2016], de origem e destino, e o terceiro para carregamento das viagens para a estrutura de grafo da biblioteca *Jung*.

Entre as unidades está o sistema de coordenadas Latitude/Longitude que representam em graus a distância de um local ao Meridiano de *Greenwich* e à Linha do Equador [PENA, 2016]. Outro sistema relevante para este trabalho é também o sistema UTM. Nele o globo é dividido em 60 fusos de 6 graus cada, em amplitude e longitude, e cada um dos fusos, também chamado de Zona UTM, a qual é numerada, inicia em 1 da esquerda para a direita, em relação à longitude 180° grau oeste [PEREIRA, 2016]. Nele, o mundo é dividido em 60 fusos, cada um de 6° de longitude. Os fusos do sistema UTM indicam em qual parte do globo está, uma vez que um par de coordenadas pode se repetir em qualquer um dos fusos [SILVA, GUALBERTO, TUPINAMBÁS, 2013]. Além desses há também o conceito de Área Heterogênea aplicado em Belo horizonte. Esse termo Área Heterogênea agrega, em média, três Setores Censitários [BHTRANS, 2012]. Adotou-se então como unidade padrão o sistema de Setor Censitário visto que o mesmo possui equivalência com os demais e que em Belo Horizonte não há como converter para outros sistemas.

As redes estudadas são representadas na forma de redes complexas por um conjunto finito $G(V, E)$ no qual um vértice $v_i$ pertencente ao conjunto $V\{v_1, v_2, ..., v_n\}$ e que representa um setor censitário. Uma aresta $e_{ij}$, que liga dois vértices $v_i$ e $v_j$ de V, pertencente ao conjunto $E\{e_1, e_2, ..., e_m\}$ e representa uma origem e destino. O peso $w_{ij}$ da aresta $e_{ij}$ representa a quantidade de viagens realizadas entre a origem $v_i$ e o destino $v_j$.

O grau do vértice $v_i$, descrito por $k_{vi}$, aqui significa o quão conectado é um setor, em outras palavras o quão relevante o mesmo é como um ponto de origem ou chegada das viagens. O grau de entrada indica a variedade do conjunto de origens de viagens para este setor, ou seja, o "interesse" de pessoas de outros setores nele. Já o grau de saída de um vértice $v_i$, representa a variedade das opões de destinos de viagens partindo de $v_i$.

Em seu trabalho, Saberi (2016) calculou uma série de propriedades no processo de caracterização das redes de Chicago e Melbourne, para possibilitar a comparação, calculamos os valores das mesmas para as redes das cidades brasileiras, são elas: o número de nós (N), o número de arestas (L), o acumulado dos pesos das arestas (T), a razão entre a quantidade de arestas e a quantidade de nós (L/N), o índice de conectividade da rede (Δ), o fluxo médio da rede (F), o grau médio (K) e do peso médio das arestas da rede (W).

O número de nós, *N*, é obtido pela contagem dos vértices, *v*, da rede. No contexto deste trabalho N é o total de setores censitários de uma cidade. De forma similar o número de arestas, *L*, é calculado a partir da contagem das arestas de uma rede de origem e destino. L representa a quantidade de origens e destinos distintas em uma cidade. O acumulado dos pesos das arestas, T, é obtido pelo somatório dos pesos das arestas da rede. No contexto deste trabalho T representa a quantidade de viagens, ou origens e destinos, que a cidade possui. O índice de conectividade da rede (Δ) indica o quão diluído está a quantidade de arestas em relação à quantidade máxima de combinações de vértices dois a dois (SABERI ET AL, 2016). O fluxo *F* é o somatório dos pesos das arestas (SABERI ET AL, 2016), e no caso de redes de transporte indica a quantidade ou recorrência de viagens de um determinado setor que em outro.

## 3. Resultados

Esta seção tem como objetivo comparar as redes de origem e destino de três grandes metrópoles brasileiras com as cidades Melbourne e Chicago. Inicialmente será feita uma comparação das propriedades das redes brasileiras, calculadas aqui, com as propriedades das redes estrangeiras, coletadas em Saberi (2016). Em seguida será feita uma análise da distribuição dos pesos das arestas, $w_{ij}$, das redes brasileiras, que no contexto deste trabalho representam a quantidade de pessoas que partem da origem *i* para o destino *j*.

### 3.1. Comparação por propriedades estruturais

A Figura 1 apresenta a comparação entre as redes das cidades brasileiras com as redes de cidades estrangeiras, os valores das propriedades podem ser visualizados na forma de radar. A escala é apresentada em logarítmico ao lado do item (a) e é idêntica para todas as figuras. As propriedades das cidades brasileiras estão representadas na cor azul e as propriedades das cidades estrangeiras em cor laranja. Em (a) as propriedades de Belo Horizonte são comparadas com Chicago e em (b) com Melbourne. No item (b) são apresentadas as propriedades de Fortaleza e Chicago, e no item (c) Fortaleza com Melbourne. Por último são comparados os dados de São Paulo com Chicago no item (e) e com Melbourne no item (f). Em resumo, Chicago tem maior semelhança com Belo Horizonte e Fortaleza e Melbourne se parece mais com São Paulo.

Quanto às propriedades de Belo Horizonte, no item (a) o número setores (N), arestas ou origens-destinos (L), viagens (T) e o peso médio são quase incidentes, o que se reflete no índice origens-destinos por setores (L/N). Já os outros valores: de grau médio (K), fluxo médio (F) e conectividade da rede (delta) estão próximos aos de Chicago, como que obedecendo uma escala ou porcentagem próxima ao valor daquela mesma propriedade em Chicago. O desenho do radar de Belo Horizonte revela-se não somente próximo, mas também em algum tipo de escala em relação ao de Chicago. Esse comportamento não se reflete em relação a Melbourne, com algumas propriedades incidentes, outras maiores e outras menores claramente não seguindo o mesmo padrão.

Destaca-se na Figura 1 (c) que Fortaleza possui o número de vértices (N) e o peso médio de aresta (W) incidentes nos valores de Chicago. Já a quantidade de arestas (L), o acumulado de pesos (T), a razão entre o número de arestas e vértices (L/N), a conectividade da rede (delta) e o grau médio (K) também se aproximam mais de Chicago. O valor mais distante é o fluxo médio (F), porém, mesmo assim o radar das propriedades também segue um padrão como que percentual em relação a Chicago. Fenômeno que não se reflete em relação a Melbourne, pelos mesmos motivos que em Belo Horizonte.

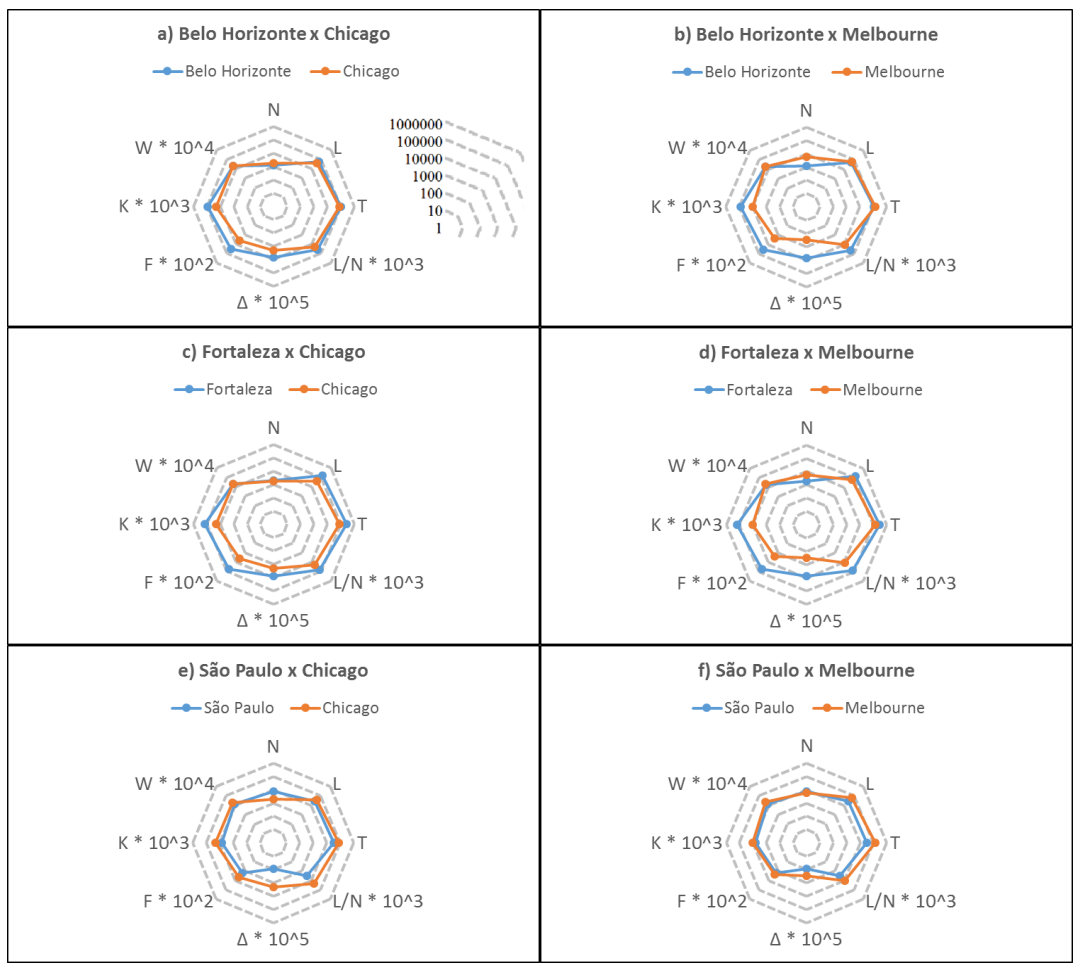

**Figura 1 – Comparativo entre das cidades brasileiras com Melbourne e Chicago.**

Pode-se constatar na Figura 1 (d) que, diferentemente de Belo Horizonte e Fortaleza, a rede de São Paulo possui várias propriedades mais parecidas com Melbourne, a começar pela quantidade de vértices (setores censitários) da rede, assim como a relação aresta por vértices (origens-destinos/setores) e a conectividade da rede. A conectividade ($\Delta$) e a relação viagens por setores (L/N), que indicam o quão diluído estão as viagens numa rede de origem-destino, em São Paulo é aproximadamente 30% de Melbourne. Essa proporção se deve ao fato da quantidade de viagens de São Paulo ser pouco mais que 30% da encontrada em Melbourne e dos pontos de origens e destinos das viagens de São Paulo ser apenas 20% maior. Ou seja, a mesma quantidade de nós e somente um terço da quantidade de viagens.

A quantidade de viagens T por sua vez interfere apenas na proporcionalidade do fluxo médio e grau médio. O fluxo médio representa a quantidade média de viagens que partem ou chegam de um setor censitário ou área heterogênea. E o grau médio indica a quantidade de destinos ou origens diferentes de viagens que partem ou chegam de setor ou área heterogênea. Nesse quesito pode-se verificar que Belo Horizonte está entre Fortaleza e Chicago, e mesmo respeitando a proporcionalidade de 3 setores censitários para cada área heterogênea, ainda está entre Fortaleza e Chicago, visto que o grau médio de Melbourne é a metade de Chicago.

O peso médio das arestas reflete um valor médio da quantidade de viagens entre uma determinada origem/destino. Exceto São Paulo, todas as outras cidades possuem um peso médio próximo a 2, ou seja, em média há pelo menos duas viagens, saindo ou chegando, executadas entre cada setor de origem e destino. Já em São Paulo essa média cai para 1,27.

### 3.2. Distribuição de peso das arestas

O peso da aresta indica o quão forte é a interação entre os nós da rede, sendo, então possível que um determinado nó, que no caso deste estudo é um setor censitário ou área heterogênea, seja muito conectado, ou seja, tenha um grau alto, mas, com peso baixo das arestas, como também tenha o grau baixo, poucas arestas, mas, com peso alto.

Para caracterização das distribuições foi aplicada uma regressão linear [REIS, 1994], que estima uma função $y = A * x^\alpha$ que descreve o valor esperado para uma variável, aplicada às distribuições de peso das arestas parte fundamental da análise. A geração das distribuições foi feita através de um programa *Java* e os gráficos foram gerados a partir das distribuições no programa *QtGrace*.

A Figura 2 ilustra as distribuições dos pesos das arestas das redes brasileiras. Foi observado a existência de leis de potência nas distribuições de Fortaleza e Belo Horizonte, tal qual foi observado em Chicago em Saberi et al (2016) e Berton (2016). Em Belo Horizonte foram encontrados expoentes com valores próximos $\alpha$ = -2.0598 e -2.2288 para 2002 e 2012 respectivamente. A semelhança dos valores de $\alpha$ sugerem que o padrão de viagens permaneceu praticamente constante durante esse período. Ainda na Figura 2, em (c), observa-se que Fortaleza também apresentou uma lei de potência na sua distribuição de pesos das arestas com valor de $\alpha$ = -2.5176.

Observa-se na Figura 2 (b) que, diferente das demais cidade brasileiras, São Paulo não confirma a presença de uma lei de potência na sua distribuição dos pesos das arestas. Os valores demasiadamente altos de $\alpha$, somado ao fato do fenômeno só está presente em duas ordens de grandeza (observar o eixo x) não revelam provas suficientes para confirmar uma relação livre de escala.

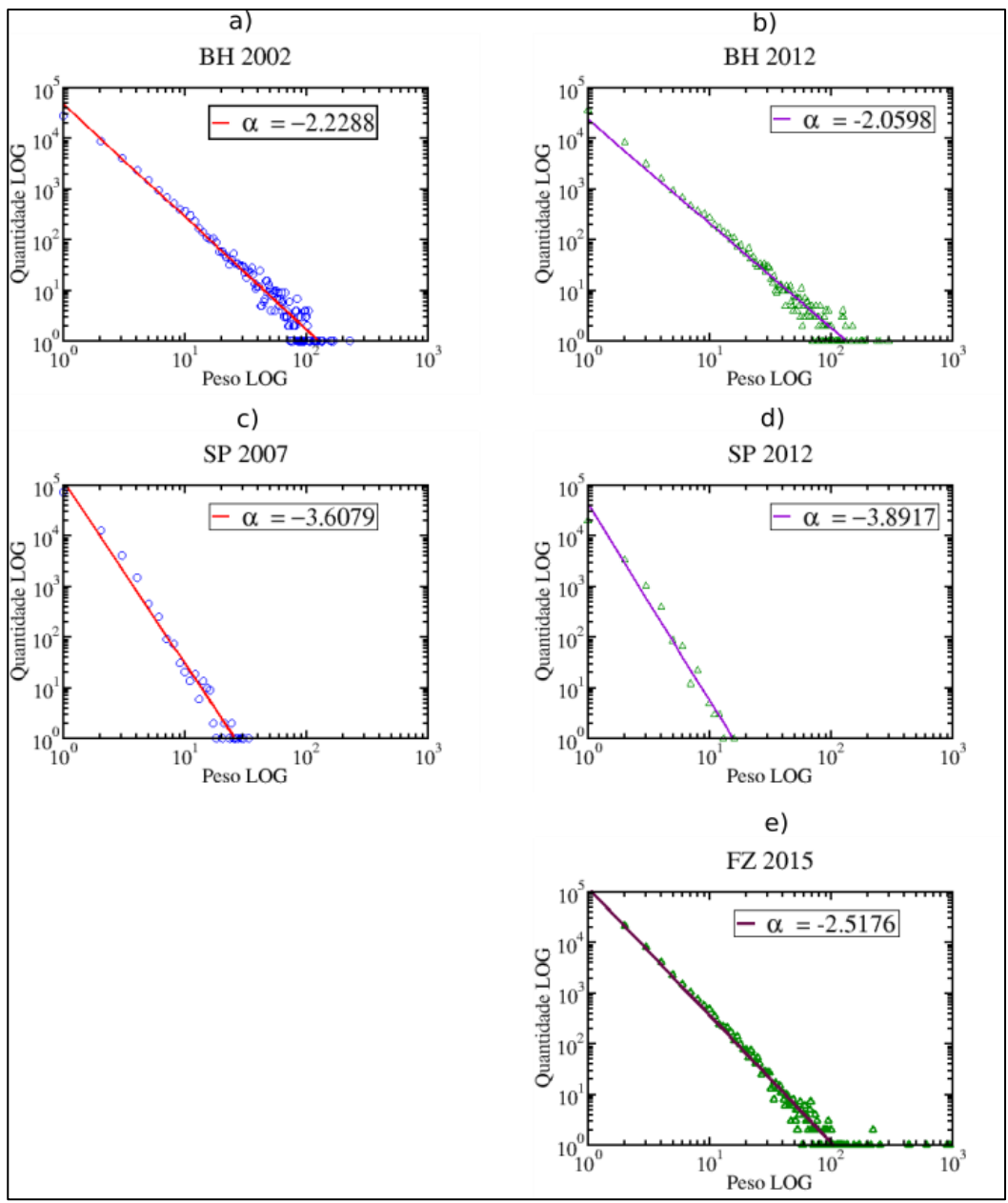

**Figura 2 – Distribuição dos pesos das arestas**

## 4. Conclusão

O principal objetivo deste trabalho é caracterizar e comparar redes de origem e destino de cidades brasileiras com cidades estrangeiras. Identificou-se a existência de dois grupos entre as cidades estudadas, o primeiro, formado por Belo Horizonte, Chicago e Fortaleza, e o segundo formado por Melbourne e São Paulo. Os grupos foram definidos tanto a partir da verificação de similaridade entre métricas de redes complexas calculadas, quanto pela verificação de similaridade a nível de lei de escala, especificamente pelas suas distribuições de pesos das arestas.

Esses resultados corroboram com o achado de Saberi (2016), que encontrou grande dissimilaridade entre as redes de origem e destino de Melbourne e Chicago. Uma das explicações que Saberi encontrou para essa diferença de características é de caráter econômico, mais precisamente devido a distribuição espacial dos polos comerciais de Chicago e Melbourne. Enquanto a cidade australiana tem diversos polos comerciais de tamanho equivalente distribuídos pelo seu espaço urbano, a cidade americana é conhecida por ter um grande polo comercial de tamanho desproporcional as demais localizações. O fato de termos encontrado similaridade entre Chicago, Fortaleza e Belo Horizonte reforça o achado de Saberi (pelo fato dessas duas cidades brasileiras serem conhecidas por também possuírem um polo comercial único em seu espaço urbano) e mais ainda, nossos resultados indicam que cidades com um único polo comercial naturalmente convergem para uma relação livre de escala na sua distribuição de viagens.

## REFERÊNCIAS BIBLIOGRÁFICAS

ANEXO 1

**Tabela I – Propriedades das redes de origem e destinos das cidades brasileiras estudadas**

| Símbolo | Unidade | Belo Horizonte | Fortaleza | Chicago | São Paulo | Melbourne |
|---|---|---|---|---|---|---|
| N | Vértices | 1255 | 2002 | 1867 | 7532 | 5998 |
| L | Arestas | 53843 | 147517 | 37527 | 25387 | 63788 |
| L/N | Arestas / Vértices | 42,902 | 73,684 | 20,1 | 3,318 | 10,63 |
| Δ | Conectividade da rede | $68,3 * 10^{-3}$ | $73,6 * 10^{-3}$ | $21,0 * 10^{-3}$ | $0,9 * 10^{-3}$ | $3,0 * 10^{-3}$ |
| T | Viagens | 108093 | 270959 | 78680 | 31854 | 133754 |
| F | Fluxo Médio | 344,5 | 541,4 | 42,1 | 16,9 | 22,3 |
| K | Grau Médio | 85,8 | 147,37 | 20,1 | 6,63 | 10,63 |
| W | Peso Médio | 2 | 1,84 | 2,1 | 1,27 | 2,1 |
| CV(F) | Coeficiente de Variação do Fluxo | 0,96 | 2,38 | 1,15 | 1,14 | 2,01 |
| CV(k) | Coeficiente de Variação do Grau | 1,04 | 1,28 | 0,93 | 1,17 | 1,4 |
| CV(w) | Coeficiente de Variação do Peso | 1,38 | 1,78 | 1,73 | 0,53 | 1,56 |